\documentclass[12pt]{iopart}

\usepackage{iopams}  
\usepackage{graphicx}
\begin{document}

\title[Quark Recombination and Diffusion]{Quark Recombination and
Heavy Quark Diffusion in Hot Nuclear Matter}

\author{R J Fries$^{1,2}$, M He$^1$, R Rapp$^1$}

\address{$^1$Cyclotron Institute and Department of Physics and Astronomy,
Texas A\&M University, College Station TX, USA}
\address{$^2$RIKEN/BNL Research Center, Brookhaven National Laboratory, Upton
  NY, USA}
\ead{rjfries@comp.tamu.edu}

\begin{abstract}
We discuss resonance recombination for quarks and show that it is compatible
with quark and hadron distributions in local thermal equilibrium. We then 
calculate realistic heavy quark phase space distributions in heavy ion
collisions using Langevin simulations with non-perturbative $T$-matrix interactions in 
hydrodynamic backgrounds. We hadronize the heavy quarks on the critical 
hypersurface given by hydrodynamics after constructing a criterion for the 
relative recombination and fragmentation contributions. We discuss the 
influence of recombination and flow on the resulting heavy meson and single 
electron $R_{AA}$ and elliptic flow. We will also comment on the effect of
diffusion of open heavy flavor mesons in the hadronic phase.
\end{abstract}

%\maketitle

Quark recombination has been identified as an important mechanism of
hadronization in high energy nuclear collisions 
\cite{Rapp:2003wn,Fries:2003vb,Fries:2003kq,Greco:2003mm,Fries:2008hs}. Evidence for 
recombination can be seen in the large baryon/meson ratios and
the quark-number scaling of elliptic flow $v_2$ at the 
Relativistic Heavy Ion Collider (RHIC) \cite{Fries:2008hs}.
First-generation recombination models that could explain these phenomena
at intermediate transverse momenta $p_T \approx 1.5 \ldots 6$ GeV/$c$ were
based on so-called instantaneous quark coalescence. They lack explicit energy 
conservation and serious concerns arise if they are applied to low $p_T$ where the 
bulk of hadrons are produced.

Hence it is important to understand how quark recombination in 
local kinetic equilibrium can be described. A promising formalism was 
put forward in \cite{Ravagli:2007xx,Ravagli:2008rt} based on a Boltzmann
equation for quarks and antiquarks scattering through resonances which
resemble mesons. This resonance recombination model (RRM) conserves
4-momentum and exhibits detailed balance. Therefore, in the long-time limit
of the Boltzmann dynamics the mesons should acquire local thermal equilibrium
with temperature and flow fields given by those of the
equilibrated quark distributions. We have numerically confirmed this
assertion both for simple blastwaves with constant hadronization
time \cite{He:2010vw}, and for the non-trivial hadronization hypersurface of a
hydrodynamic evolution of the quark phase \cite{He:2011qa}.
Hence resonance recombination preserves local thermal equilibrium for arbitrary
flow fields and for any realistic hypersurface.
An important corollary is the fact that any quark-number or kinetic energy
scaling at \emph{low} hadron-$p_T$ is not due to quark recombination. 
Local equilibration would not allow any microscopic information about
hadronization to be detected, and we explicitely show in
Ref.\ \cite{He:2010vw}  how a realistic flow field together with a reasonable 
freeze-out hierarchy for hadrons can mimic quark number- and kinetic energy 
scaling at RHIC at \emph{low} $p_T$ to good accuracy.

Heavy charm ($c$) and bottom ($b$) quarks have been widely discussed as ideal 
internal probes for quark gluon plasma, see \cite{Rapp:2009my} and references
therein. The large masses increase their thermalization times, and the
degree of thermalization in heavy ion collisions encodes valuable information
about their interactions with quark gluon plasma (QGP). Here we report on our 
effort to set up a
formalism that is consistently based on the notion of strongly coupled QGP,
both for the diffusion of heavy quarks through QGP, and their hadronization
through an appropriate superposition of recombination and
fragmentation \cite{He:2011qa}. The latter provides an
opportunity to extend the resonance recombination formalism to quark
distributions away from local thermal equilibrium. 

The dynamics of the heavy quarks in a background medium can be described by a 
relativistic implementation of the Langevin equation \cite{Hanggi2009}.
We use the ideal, 2+1-D, boost-invariant hydro code AZHYDRO 
\cite{Kolb:2003dz,KSH:2000} to
model the locally equilibrated background QGP medium.
The drag and diffusion coefficients of the heavy quarks in the medium are
calculated from non-perturbative $T$-matrix results on heavy-light quark
interactions \cite{Riek:2010fk,Riek:2010py}. The $T$-matrix approach exhibits 
Feshbach resonances both in the color-singlet and color-triplet channel and
leads to resonant interactions up to $\approx 1.5 \, T_c$. The resonant
relaxation rates are substantially enhanced compared to the corresponding 
perturbative elastic rates. The initial heavy quark momentum spectra are taken
from perturbative calculations cross-checked with measured $p+p$ semi-leptonic
decay spectra \cite{vanHees:2005wb}, while the spatial distribution is
determined by the binary collision density.

\begin{figure}
\centerline{\includegraphics[width=7.5cm]{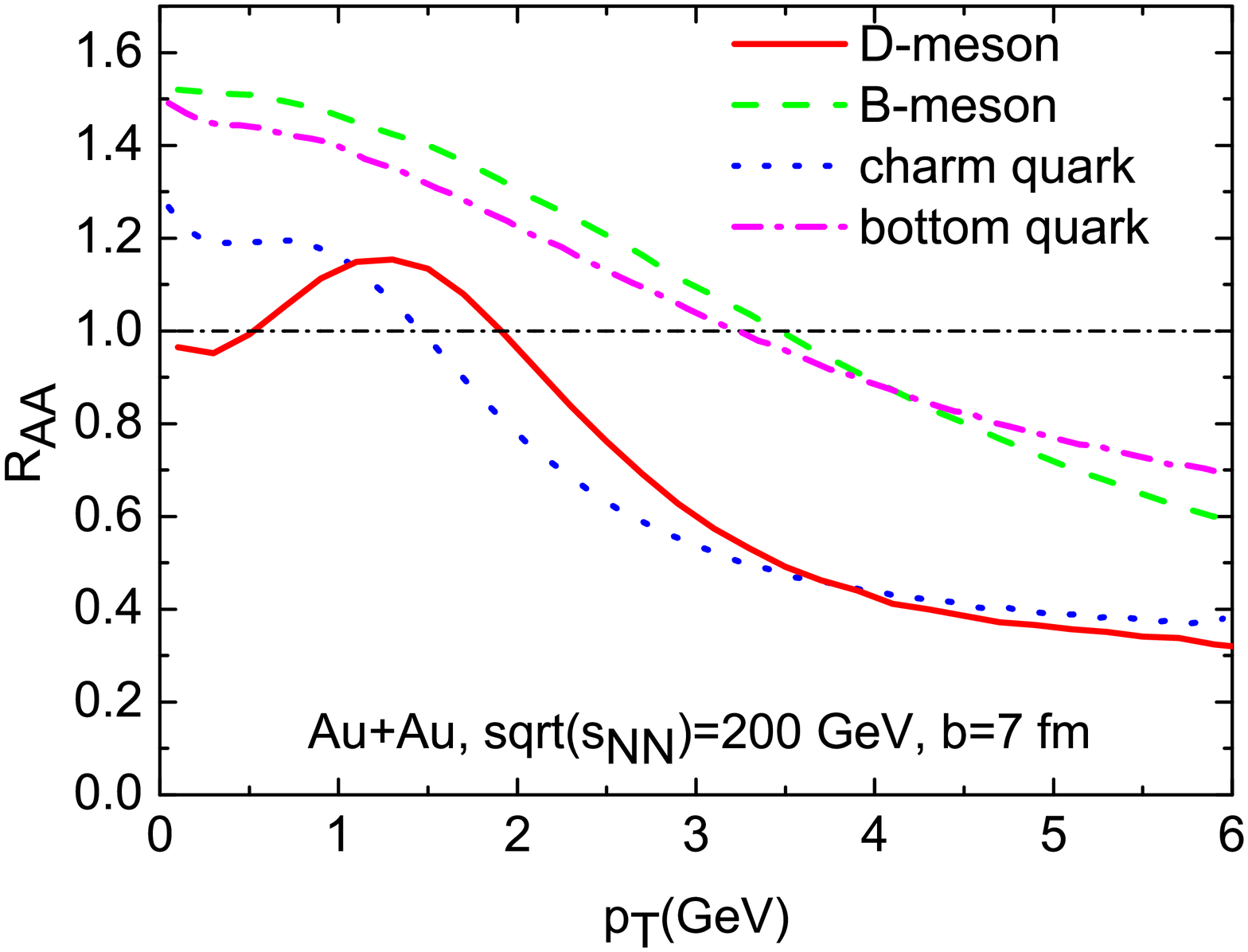}}
\caption{$D$- and $B$-meson nuclear modification factors for semi-central
  Au+Au collisions at RHIC compared to the $c$ and $b$ quark $R_{AA}$ from 
  which they originate.}
\label{fig:1}
\end{figure}

We employ a test particle method for our Langevin simulation. We have checked
that in the limit of very large (unphysical) relaxation rates our charm quark
distributions approach the equilibrium given by the hydrodynamic background.
Figure \ref{fig:1} shows the nuclear modification factor $R_{AA}$ for $c$ and
$b$ quarks on the hadronization hypersurface (given by AZHYDRO) calculated
from our Langevin with non-perturbative $T$-matrix coefficients for
semi-central collisions at RHIC. We observe
about 60\% quenching for charm quarks at large momenta $p_t$ which is
reflected in an excess at small $p_t$ due to conservation of heavy quarks.
Suppression for $b$ is significantly smaller.

Hadronization of heavy quarks should proceed through coalescence with light
quarks from the medium if heavy-light scattering rates are large enough to
allow for such interactions in the color-singlet channel during the pertinent time
interval (e.g.\ the duration of the mixed phase). On the other hand, if such
heavy-light interactions are rare, e.g.\ at large $p_t$, independent fragmentation
of the heavy quark should occur.  For a given heavy quark momentum in its
local fluid rest frame we estimate the scattering rate $\Gamma_Q^{\mathrm{res}}$
in the color singlet channel in the $T$-matrix approach at $T_c$. We approximate
the coalescence probability through $P_{\mathrm{coal}}(p) = \Delta
\tau_{\mathrm{res}} \Gamma_Q^{\mathrm{res}}$ (or $P_{\mathrm{coal}} =1$ if the
product exceeds one). In practice we boost the rates into the lab frame and apply them
as a function of heavy quark $p_t$ with $\Delta \tau_{\mathrm{res}} = 2$
fm/$c$ (motivated by the fact that this time interval is well below the
duration of the mixed phase in AZHYDRO) which
leads to  $P_{\mathrm{coal}} \to 1$ at vanishing $p_t$.
Note that this approach treats the interactions of heavy
quarks in the medium and their hadronization using the same non-pertrubative
dynamics. It also ensures that $P_{\mathrm{coal}} \to 0$ for very large $p_t$.
We evaluate $P_{\mathrm{coal}}$ for each test particle and apply fragmentation
or resonance recombination accordingly.

\begin{figure}
\centerline{\includegraphics[width=7.5cm]{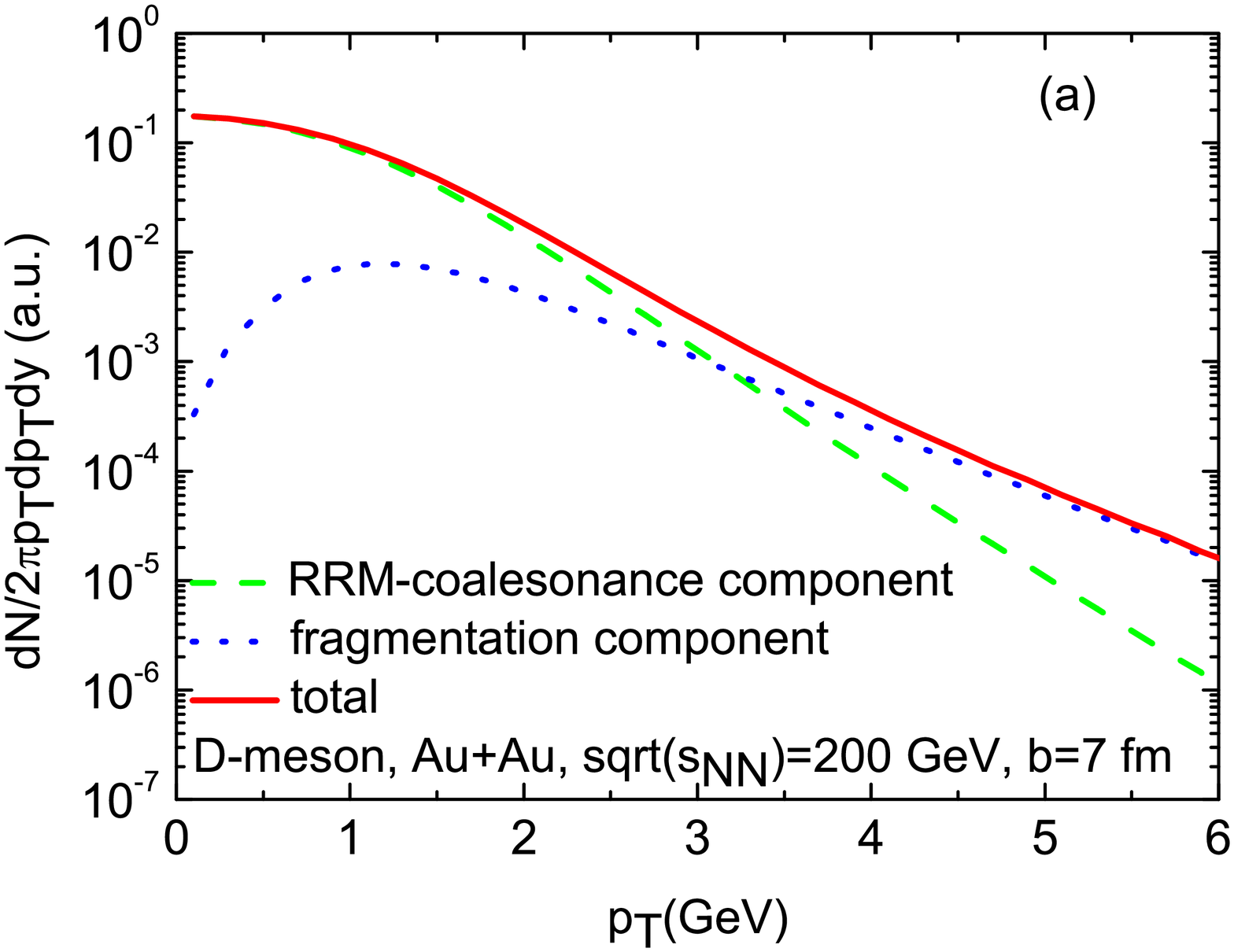}
\includegraphics[width=7.5cm]{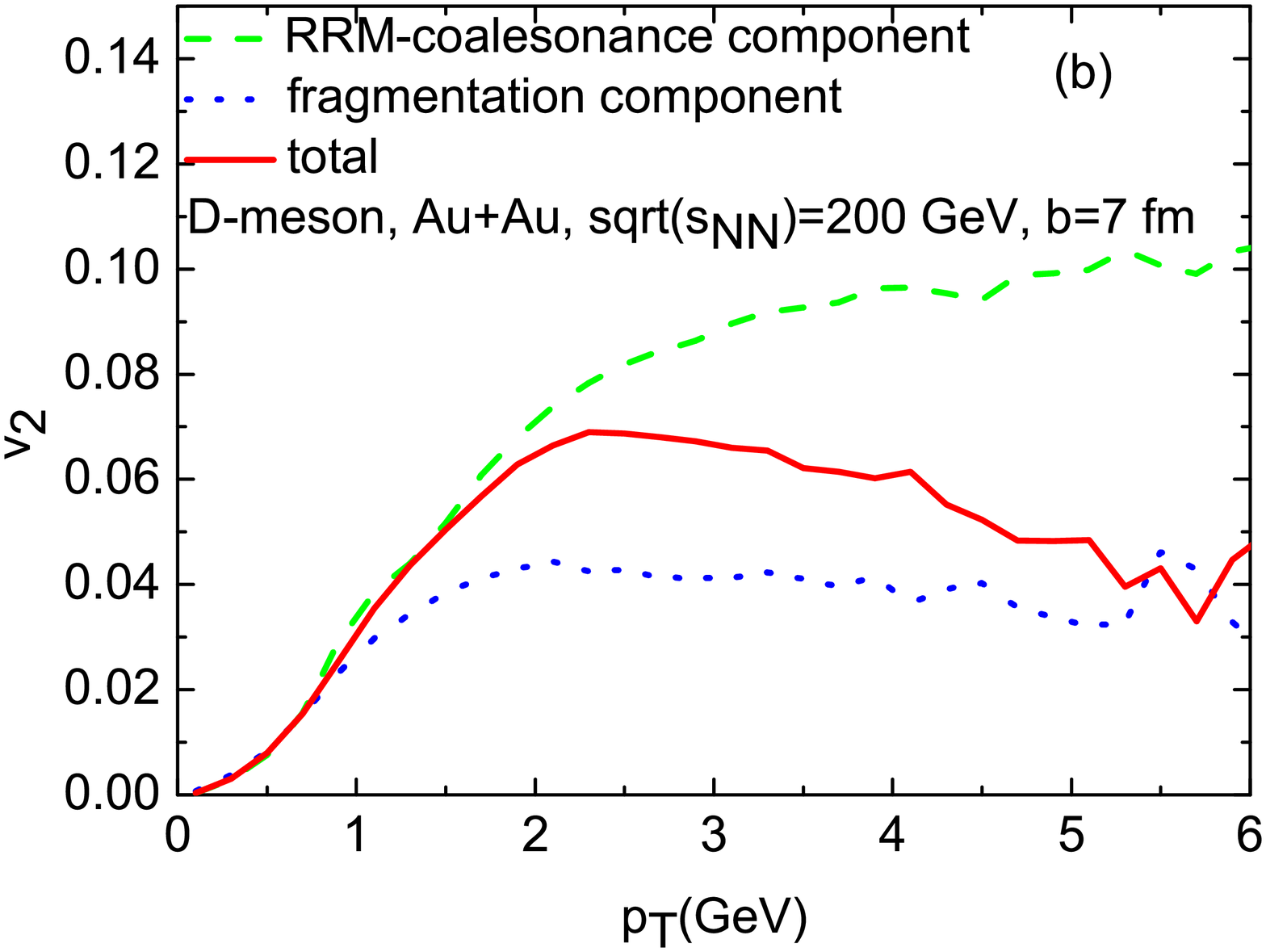}}
\caption{(a) The coalescence, fragmentation and total $D$-meson
  $p_T$-spectrum for semi-central Au+Au collisions at top RHIC energy. The
total spectrum is normalized to one test-particle. (b): The same three
contributions to the elliptic flow $v_2$.}
\label{fig:2}
\end{figure}

\begin{figure}
\centerline{\includegraphics[width=7.5cm]{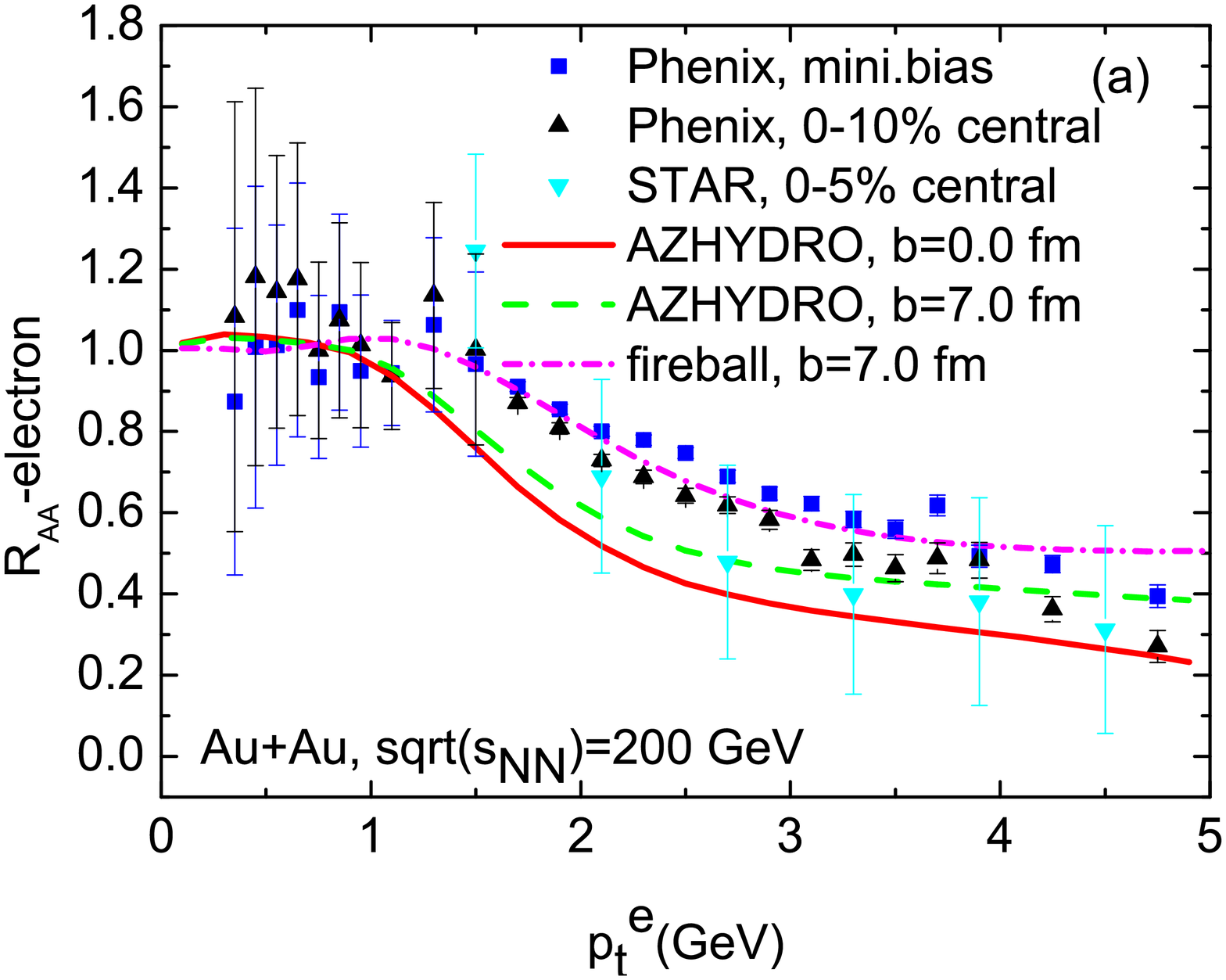}
\includegraphics[width=7.5cm]{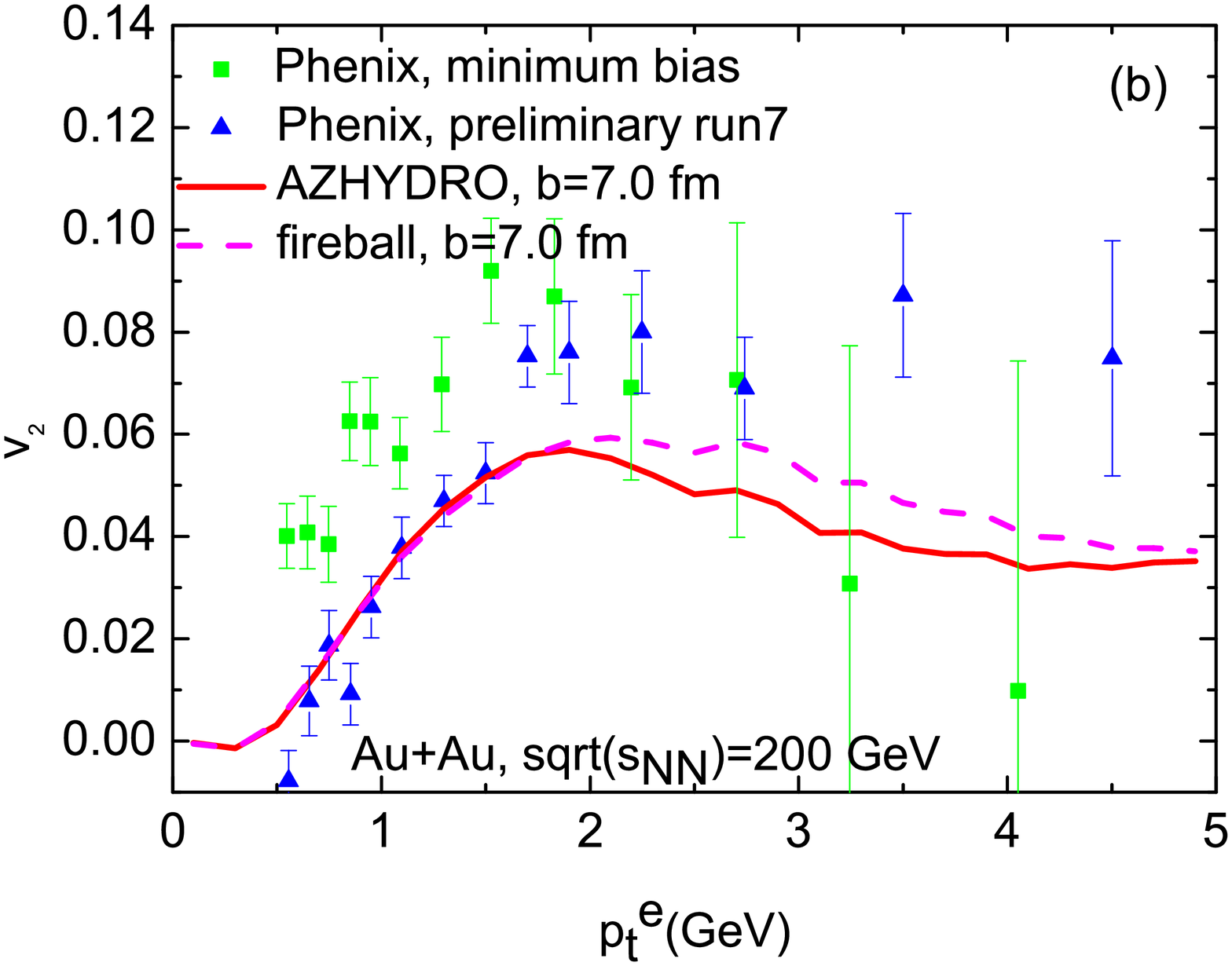}}
\caption{(a) Electron $R_{AA}$ from semi-leptonic decays of $D$ and $B$ mesons
  for central ($b=0$ fm) and semi-central ($b=7$ fm) Au+Au collisions at RHIC.
  using either AZHYDRO or a parameterized fireball as the background medium.
  We compare to data from PHENIX \cite{Adare:2006nq} and STAR \cite{Abelev:2006db}.
  (b) Electron $v_2$ from the same source.}
\label{fig:3}
\end{figure}

Figure \ref{fig:2} shows spectra and elliptic flow $v_2$ for $D$ mesons for
semi-central collisions at RHIC. Fragmentation and recombination contributions
are shown together with the total. Figure \ref{fig:1} provides $R_{AA}$ for
both $D$ and $B$ mesons. We notice that resonance recombination provides
a significant enhancement of both radial and elliptic flow. The former can be
seen from the prominent flow ``bump'' developing in $R_{AA}$ for $D$ mesons. 
Resonance recombination acts as an additional interaction of heavy
quarks with the medium, making a noticeable contribution to further
equilibration. In Figure \ref{fig:3} we display $R_{AA}$ and $v_2$ of electrons
from semi-leptonic decays of $D$ and $B$ mesons, compared to experimental
data. The comparison with data is favorable if one keeps in mind two facts: 
AZHYDRO is tuned to kinetic freeze-out and exhibits too little radial flow
around $T_c$. This becomes obvious if the results with AZHYDRO are compared to 
those using a fireball parameterization with the correct radial flow at
$T_c$ (see \cite{He:2010vw,He:2011qa} for details). Secondly, we expect an
increase in elliptic flow of about 20-30\% in the hadronic phase. We have
recently calculated relaxation rates of charm in a hadron gas and have
found that at $T_c$ the results tend to be close to those of the $T$-matrix
calculation in QGP \cite{He:2011yi}. We will elaborate on both points further
in a forthcoming publication.

This work was supported
by the U.S.\ National Science Foundation (NSF) through CAREER grant PHY-0847538
and grant PHY-0969394, by the A.-v.-Humboldt
Foundation, by the RIKEN/BNL Research Center and DOE grant
DE-AC02-98CH10886, and by the JET Collaboration and
DOE grant DE-FG02-10ER41682.

\end{document}